\documentclass[11pt,a4paper,english,showpacs,superscriptaddress,aps]{revtex4-1}
\usepackage{amssymb}
\usepackage{amsmath}
\usepackage{graphicx}
\usepackage{babel}
\usepackage{hyperref}

\begin{document}

\title{Noncommutative supersymmetric Chern--Simons-matter model}

\author{A.~C.~Lehum}
\email{andrelehum@ect.ufrn.br}
\affiliation{Escola de Ci\^encias e Tecnologia, Universidade Federal do Rio Grande do Norte\\
Caixa Postal 1524, 59072-970, Natal, RN, Brazil}


\begin{abstract}
Using the superfield formalism and implementing the canonical noncommutativity, the K\"{a}hlerian effective superpotential is evaluated in the three-dimensional noncommutative supersymmetric Chern--Simons-matter model at the two-loop order. The computation of the K\"{a}hlerian effective superpotential is enough to determine whether the model can exhibit spontaneous (super) symmetry breaking. It is shown that the model possesses a spontaneous gauge symmetry broken phase, generating masses for the scalar and gauge superfields at the two-loop order. Just as for the commutative version, in the noncommutative case, the supersymmetry cannot be broken by radiative corrections via the Coleman--Weinberg mechanism.  
\end{abstract}



\maketitle

\section{Introduction}

In the 40s  of the last century, Heisenberg suggested that an uncertainty principle in space--time coordinates should improve the ultraviolet behaviour of quantum field theories. Inspired by this idea, the first paper on noncommutative field theory (NCFT) was published in 1947~\cite{Snyder:1947nq}, but due to the success of renormalization theory, this idea was forgotten until the 90s. We can say that there are two facts responsible for the increasing interest in such theories. The first is related to the discovery that the noncommutative Yang--Mills theory arises as a low energy limit of a string theory~\cite{Seiberg:1999vs}. The second motivation is related to "space--time foam," i.e., the idea that at the Planck length order ($10^{-33}$~cm), space--time loses its continuum structure and should involve quantum fluctuations of topology and geometry~\cite{Doplicher:1994tu}. The formulation of an NCFT would be a simple way to implement these ideas.

There are several ways to implement the noncommutativity of space--time coordinates in a field theory, but apparently all of them share one remarkable characteristic, the so-called UV/IR mixing~\cite{Minwalla:1999px}, that is, a transmutation of the original ultraviolet (UV) divergence in the ordinary theory to an infrared (IR) divergent behaviour in its noncommutative extension. This dangerous UV/IR mixing can invalidate the perturbative expansion. A way to avoid this issue is to work with less UV-divergent theories, suggesting supersymmetric models. It is well-known that supersymmetry improves the ultraviolet behaviour of the models, and in many cases, makes the theories finite (see, e.g., Refs. \cite{N4S,n2sc,RuizRuiz:1997jq,Ferrari:2007mh}). This improvement is due to cancellations between bosonic and fermionic parts of higher order divergences present in a supergraph. The supersymmetric noncommutative models are less susceptible to have UV/IR mixing, being natural candidates for a consistent NCFT~\cite{Girotti:2000gc}. 

The noncommutativity of space--time coordinates can be expressed by
\begin{eqnarray}\label{in01}
[x^\mu,x^\nu]=i\Theta^{\mu\nu}~,
\end{eqnarray}

\noindent
where $\Theta^{\mu\nu}$ is an antisymmetric constant (canonical noncommutativity) matrix, which is suggested to be of the order of ${l}_P^2$, with ${l}_P$ the Planck length. In contrast to a constant matrix, one could consider $\Theta^{\mu\nu}$ as an independent quantity having a canonical conjugate momentum, see for instance Refs.~\cite{Abreu:2010mt,Amorim:2010qj}, or a dynamical noncommutativity as discussed in Refs.~\cite{Gomes:2009rz}.

We can implement the noncommutativity to a field theory replacing the ordinary product by the Moyal one, denoted by a $*$, where the Moyal product between two fields is given by
\begin{eqnarray}\label{in02}
\phi_1(x)*\phi_2(x)=\phi_1(x) \exp\left[ -\frac{i}{2}\overleftarrow{\partial_\mu}\Theta^{\mu\nu}\overrightarrow{\partial_\nu} \right]\phi_2(x)~,
\end{eqnarray}

\noindent
which has the important property
\begin{eqnarray}\label{in03}
\int{d^Dx}~\phi_1(x)*\phi_2(x)*\cdots*\phi_n(x)=\int{d^Dx}~\phi_1(x)\phi_2(x)*\cdots*\phi_n(x)~,
\end{eqnarray}

\noindent
from which we can see that, in particular, the kinetic part of an action is unaffected. Therefore, in this approach, all information about the noncommutativity of space--time coordinates comes from the interaction terms.

Gauge theories are of great interest in physics, and noncommutative extensions of ordinary gauge theories were widely studied in several aspects, both in four~\cite{Matusis:2000jf,Buchbinder:2001hn,Bichl:2002wb,Ferrari:2004ex,Ishihara:2012ha} and lower dimensions of space--time~\cite{SheikhJabbari:2001au,Das:2001kf,Ferrari:2003ma,Ferrari:2006xx,Armoni:2011pa,Gomes:2011aa}. In particular, one aspect which has not been contemplated in earlier works is the generation of mass by radiative corrections in three dimensions. Recently, three-dimensional supersymmetric gauge models have attracted some attention because they are candidates for describing M2 branes~\cite{Bagger:2006sk,Gustavsson:2007vu}, in particular several aspects of supersymmetric Chern--Simons-matter models (SCSM) have been studied~\cite{Gaiotto:2007qi,Ferrari:2010ex,Lehum:2010tt,Gallegos:2011ux}. 

In this work we investigate some perturbative aspects of the noncommutative ${\cal N} = 1$ supersymmetric Chern--Simons-matter model (NCSCSM) in three-dimensional space--time. We have used the superfield formalism because it is a more convenient way to perform Feynman graphs in supersymmetric theories. It keeps the supersymmetry manifest in all stages of the calculations, avoiding potential problems in the renormalization procedure, e.g., the lacking of a supersymmetric renormalization presented in Ref.~\cite{Inami:2000eb} is not a problem when supergraph techniques are used~\cite{Ferrari:2006xx}. 

This article is organized as follows. In Sec.~\ref{nccsmm}, we present the model and compute the propagators in a convenient approximation. In Sec.~\ref{eoes}, we evaluate the K\"{a}hlerian effective superpotential up to two-loop order, studying its vacuum properties. In Sec.~\ref{fr}, we present our final remarks.

\section{Noncommutative Supersymmetric Chern--Simons-matter model}\label{nccsmm}

The NCSCSM is defined through the action
\begin{eqnarray}\label{eq1}
S&=&\int{d^5z}\Big{\{}-\frac{1}{2}\Gamma^{\alpha}*W_{\alpha}
-\frac{ig}{12}\{\Gamma^{\alpha},\Gamma^{\beta}\}_{*}* D_{\beta}\Gamma_{\alpha}
-\frac{g^2}{24}\{\Gamma^{\alpha},\Gamma^{\beta}\}_{*}*\{\Gamma_{\alpha},\Gamma_{\beta}\}_{*}\nonumber\\
&&-\frac{1}{2}{\nabla^{\alpha}\bar\Phi}*\nabla_{\alpha}\Phi -\lambda(\bar\Phi*\Phi)^2_*+GF+FP\Big{\}}~,
\end{eqnarray}

\noindent
where $\Gamma_\alpha$ is the gauge superpotential,  $\nabla^{\alpha}=(D^{\alpha}-ig\Gamma^{\alpha})$ is the gauge supercovariant derivative, $D_{\alpha} = \partial_{\alpha} + i \theta^{\beta} \partial_{\alpha \beta}$ is the supersymmetric covariant derivative, and $W^{\alpha}$ is the covariant field strength given by
\begin{eqnarray}\label{eq1a}
W_{\alpha}&=&\frac{1}{2}D^{\beta}D_{\alpha}\Gamma_{\beta}-\frac{ig}{2}[\Gamma^{\beta},D_{\beta}\Gamma_{\alpha}]_*
-\frac{g^2}{6}[\Gamma^{\beta},\{\Gamma_\beta,\Gamma_\alpha\}_*]_*~.\nonumber
\end{eqnarray}

\noindent
The signature is $(-,+,+)$, and we are using the notations and conventions of Ref.~\cite{Gates:1983nr}.

This model exhibits spontaneous gauge symmetry breaking in the presence of a mass term to the scalar superfield~\cite{Lehum:2007nf}. Without a mass term $\int{d^5z}m\bar\Phi*\Phi$, the model defined by Eq.(\ref{eq1}) does not exhibit spontaneous gauge symmetry (nor supersymmetry) breaking at the classical level. To verify if quantum corrections can change this feature, it is enough to evaluate the effective K\"{a}hlerian superpotential~\cite{Burgess:1983nu,Ferrari:2010ex,Lehum:2010tt,Gallegos:2011ux}. To do this, let us dislocate the scalar superfields $\Phi$ and $\bar{\Phi}$ by the constant classical superfield $\varphi=\sigma_1-\theta^2\sigma_2$ as follows   
\begin{eqnarray}\label{eq2}
\Phi&\rightarrow & \frac{1}{\sqrt{2}}(\Phi_1+\varphi+i\Phi_2)~,\nonumber\\
\Phi&\rightarrow & \frac{1}{\sqrt{2}}(\Phi_1+\varphi-i\Phi_2)~,
\end{eqnarray}

\noindent
where we assume $\langle\Phi\rangle=\langle\bar{\Phi}\rangle=\dfrac{\varphi}{\sqrt{2}}$ and $\langle\Phi_1\rangle=\langle\Phi_2\rangle=0$ in all orders in perturbation theory.

Rewriting the action (\ref{eq1}) in terms of real quantum superfields $\Phi_1$ and $\Phi_2$ using the above statement, we obtain
\begin{eqnarray}\label{eq3}
S&=&\int{d^5z}\Big{\{}-\frac{1}{2}\Gamma^{\alpha}*W_{\alpha}
-\frac{ig}{12}\{\Gamma^{\alpha},\Gamma^{\beta}\}_{*}* D_{\beta}\Gamma_{\alpha}
-\frac{g^2}{24}\{\Gamma^{\alpha},\Gamma^{\beta}\}_{*}*\{\Gamma_{\alpha},\Gamma_{\beta}\}_{*}\nonumber\\
&&+\frac{1}{2} \Phi_1(D^2-3\lambda\varphi^2)\Phi_1 
+\frac{1}{2} \Phi_2(D^2-\lambda\varphi^2)\Phi_2
+D^2\varphi\Phi_1
+\frac{1}{2} \varphi D^2\varphi\nonumber\\
&&+i\frac{g}{4}\Big([\Phi_1,D^{\alpha}\Phi_1]_**\Gamma_{\alpha}
+[\Phi_2,D^{\alpha}\Phi_2]_**\Gamma_{\alpha}
+i\{\Phi_2,D^{\alpha}\Phi_1\}_**\Gamma_{\alpha}\nonumber\\
&&-i\{\Phi_1,D^{\alpha}\Phi_2\}_**\Gamma_{\alpha}
+2iD^{\alpha}\varphi\Gamma_{\alpha}\Phi_2-2i\varphi D^{\alpha}\Phi_2\Gamma_\alpha\Big)
-\frac{g^2}{4}\varphi^2\Gamma^{\alpha}*\Gamma_{\alpha}
\nonumber\\
&&-\frac{g^2}{2}\varphi\Phi_1*\Gamma^{\alpha}*\Gamma_{\alpha}
-\frac{g^2}{4}\left(\Phi_1*\Phi_1+\Phi_2*\Phi_2+i[\Phi_1,\Phi_2]_*\right)*\Gamma^{\alpha}*\Gamma_{\alpha}\nonumber\\
&&-\frac{\lambda}{4}(\Phi_1*\Phi_1)^2_*-\frac{\lambda}{4}(\Phi_2*\Phi_2)^2_*
-\lambda\Phi_1*\Phi_1*\Phi_2*\Phi_2
+\frac{\lambda}{2}(\Phi_1*\Phi_2)_*^2\nonumber\\
&&-\lambda\varphi\Phi_1*(\Phi_1*\Phi_1+\Phi_2*\Phi_2)
-\lambda\varphi^3\Phi_1-\frac{\lambda}{4}\varphi^4
+\frac{1}{2\xi}\left(D^{\alpha}\Gamma_\alpha+\frac{\xi}{2}g\varphi\Phi_2\right)^2\nonumber\\
&&+ \bar{C}\left(D^2+\frac{\xi}{4}g^2\varphi^2\right)C+\frac{\xi}{8}g^2\varphi\bar{C}*\{ \Phi_1,C\}_* -i\frac{\xi}{8}g^2\varphi\bar{C}*[\Phi_2,C]_*\Big{\}}~,
\end{eqnarray}

\noindent
where the last line is the Fadeev--Popov term related to the $R_\xi$ gauge-fixing.

The quadratic part of the action for the quantum superfields can be written as 
\begin{eqnarray}\label{eq3a}
S_{2}&=&\int{d^5z}\Big{\{} -\frac{1}{2}\Gamma^{\alpha}*W_{\alpha} -\frac{g^2}{4}\varphi^2 \Gamma^{\alpha}*\Gamma_{\alpha} 
+\frac{1}{2\xi}\left(D^{\alpha}\Gamma_\alpha\right)^2\nonumber\\ &&+\frac{1}{2} \Phi_1(D^2-3\lambda\varphi^2)\Phi_1 +\frac{1}{2} \Phi_2\left[D^2-\left(\lambda-\dfrac{\xi}{4}g^2\right)\varphi^2\right]\Phi_2\nonumber\\ &&+\bar{C}\left(D^2+\frac{\xi}{4}g^2\varphi^2\right)C+(\textrm{interaction~terms})\Big{\}}~, 
\end{eqnarray}

\noindent from which the free propagators of the interacting fields of the model are derived as
\begin{eqnarray}\label{props}
\langle T~\Phi_1(k,\theta)\Phi_1(-k,\theta')\rangle &=&-i\frac{D^2-M_{1}}{k^2+M_{1}^2}\delta^{(2)}(\theta-\theta')~,\nonumber\\
\langle T~\Phi_2(k,\theta)\Phi_2(-k,\theta')\rangle&=&-i\frac{D^2-M_{2}}{k^2+M_{2}^2}\delta^{(2)}(\theta-\theta')~,\nonumber\\
\langle T~\Gamma_{\alpha}(k,\theta)\Gamma_{\beta}(-k,\theta')\rangle&=&-\frac{i}{2}
\Big[\frac{(D^2-M_{A})D^2D_{\beta}D_{\alpha}}{ k^2(k^2+M_{A}^2)}\nonumber\\
&-&\xi\frac{(D^2-\xi M_{A})D^2D_{\alpha}D_{\beta}}
{k^2(k^2+\xi^2M_{A}^2)}\Big]\delta^{(2)}(\theta-\theta')~,
\end{eqnarray}

\noindent where in the supersymmetric Landau gauge $\xi=0$, the ``masses''~are
\begin{eqnarray}\label{masses}
\quad M_{1}=3\lambda \varphi^2, \quad M_{A}=\frac{g^2\varphi^2}{2}, \quad M_{2}=\lambda \varphi^2~.
\end{eqnarray}

It is well-known that the effective potential is a gauge-dependent quantity, as discussed by Jackiw in Ref.~\cite{Jackiw:1974cv}. We have chosen to work in the supersymmetric Landau gauge for simplicity. 

The physical masses are obtained from (\ref{masses}) evaluating them in the minimum of the effective potential which will be calculated in the next section. Note that $M_2$ is a gauge-dependent quantity, Eq.(\ref{eq3a}), just as in the usual Higgs model. Therefore $\Phi_2$ becomes a nonphysical degree of freedom. This degree of freedom was absorbed by the gauge superfield, appearing as a massive pole in the gauge superfield propagator, Eq.(\ref{props}).

\section{Evaluation of the effective superpotential}\label{eoes}

The effective potential is an important approach to understand the quantum behaviour of physical systems through classical concepts, being a very natural way to argue about spontaneous symmetry breaking. In particular, for supersymmetric theories, it is enough to compute the K\"ahlerian effective superpotential to see whether a model is passive in exhibiting spontaneous (super) symmetry breaking~\cite{Ferrari:2010ex,Burgess:1983nu}. 

In the K\"ahlerian approximation~\cite{Ferrari:2009zx}, the classical effective action is
\begin{eqnarray}\label{keff00}
\Gamma^{(0)}&=&-\int{d^5z}~\frac{\lambda}{4}\varphi^4~.
\end{eqnarray}
 
The one-loop contribution to the noncommutative effective K\"{a}hlerian superpotential is just the trace of the superdeterminant, which is given by
\begin{eqnarray}\label{keff01}
\Gamma^{(1)}&=&\dfrac{i}{2}\mathrm{Tr}\ln[D^2+M_1]
+\dfrac{i}{2}\mathrm{Tr}\ln[D^2+M_2]
+\dfrac{i}{2}\mathrm{Tr}\ln\left[-\dfrac{i}{2}{\partial^{\beta}}_{\alpha}
+\dfrac{{C^{\beta}}_{\alpha}}{2}D^2+{C^{\beta}}_{\alpha}M_A\right].
\end{eqnarray}

\noindent
Proceeding as in Ref.~\cite{Ferrari:2009zx}, the one-loop contribution to the effective action is
\begin{eqnarray}\label{keff02}
\Gamma^{(1)}&=&\dfrac{1}{16\pi}\int{d^5z}\Big{\{}
10\lambda^2\varphi^4+\dfrac{g^4\varphi^4}{4}
~\Big\}.
\end{eqnarray}

Up to the one-loop order, there is no change in the phase structure of this model, moreover up to this order the noncommutativity of space--time has no influence over the superpotential. The two-loop diagrams have logarithmic UV divergences, and it is known that, from commutative cases~\cite{Tan:1996kz,Tan:1997ew,Dias:2003pw,Ferrari:2010ex,Lehum:2010tt,Gallegos:2011ux}, at this order there is a modification in the phase structure of the model. 

Evaluating the two-loop diagrams depicted in Figs. (\ref{Fig1}) and (\ref{Fig2}) (see Appendix \ref{twoloop} for details) and considering the small noncommutativity limit $\Theta\ll1$ (actually suggesting $\Theta\sim l_P^2$, with $l_P$ being the Plank length), 
after summing up all contributions (i.e., tree, one-loop and two-loop contributions), the noncommutative K\"ahlerian effective superpotential, $\Gamma=-\int{d^5z}K_{eff}$, can be written as  
\begin{eqnarray}\label{keff03}
K_{eff}=\left[\frac{\lambda}{4}+f(\lambda,g,\epsilon)\right]\varphi^4+e\varphi^4\ln{\frac{\varphi^2}{\mu}}+\frac{h(g,\lambda)}{\Theta^2\varphi^4}+C\varphi^4+\mathcal{O}(\Theta)~,
\end{eqnarray}

\noindent
where $\mu$ is a mass scale introduced by the renormalization by dimensional reduction~\cite{Siegel:1979wq}, $C$ is a counterterm, $f(\lambda,g,1/\epsilon)$ is a function of the coupling constants, and $\epsilon=(D-3)$, $h(g,\lambda)$ is a function of the coupling constants, $e=a_1~g^6+a_2~g^4\lambda+a_2~g^4\lambda+a_3~g^2\lambda^2+a_4~\lambda^3$, with the $a_i$ numerical factors.

The UV divergences expressed in terms of $1/\epsilon$ appearing in the above equation can be removed through the following renormalization condition:
\begin{eqnarray}\label{keff04}
\frac{\partial K_{eff}}{\partial\varphi}\Big{|}_{\varphi=v}=0~,
\end{eqnarray}

\noindent
where $v$ is the renormalization mass scale. Such a condition is equivalent to imposing the vanishing of the tadpole equation.  

Solving Eq.(\ref{keff04}) for $C$ and substituting it back into Eq.(\ref{keff03}), we obtain the following renormalized noncommutative K\"ahlerian effective superpotential:
\begin{eqnarray}\label{keff05}
K_{eff}=e\varphi^4\left(\frac{1}{2}-\ln{\frac{\varphi^2}{v^2}}\right)+\frac{h(\lambda,g)(\varphi^8+v^8)}{\Theta^2v^8\varphi^4}+\mathcal{O}(\Theta)~,
\end{eqnarray}

\noindent
which obviously has a minimum at $\varphi=\pm v$ due to Eq.(\ref{keff04}).

The existence of a minimum for the K\"ahlerian superpotential is enough to ensure that this model does not exhibit spontaneous supersymmetry breaking by the Coleman--Weinberg mechanism~\cite{Ferrari:2010ex,Lehum:2010tt,Gallegos:2011ux}. Once the minimum of $K_{eff}$ is located at $\varphi=\pm v$, we observe a spontaneous generation of mass in the supersymmetric phase for matter and gauge superfields, $M_{1}=3\lambda v^2$ and $M_{A}=\dfrac{g^2v^2}{2}$, with the mass ratio $\dfrac{M_1}{M_A}=\dfrac{6\lambda}{g^2}$. The masses $M_{1}$ and $M_{A}$ are obtained from Eq.(\ref{masses}), computing them in the minimum of the $K_{eff}$, i.e., $|\varphi|=v$. As usual in the Higgs mechanism, the Goldstone (super) boson becomes a fictitious field and its degree of freedom is absorbed by the gauge superfield, due to generation of mass.    

An interesting feature is the presence of a singularity in the limit $\Theta\rightarrow0$, with which the commutative limit of such a model breaks up. Such a singularity, caused by a UV/IR mixing present in the vacuum diagrams, also appears for the Wess--Zumino model discussed in Ref.~\cite{Ferrari:2009zx}.

\section{Concluding remarks}\label{fr}

In this work, using the superfield formalism, we investigated some perturbative aspects of the noncommutative supersymmetric Chern--Simons-matter model (NCSCSM) in three space--time dimensions. We computed the noncommutative K\"alerian effective superpotential in the small noncommutativity limit, i.e., $\Theta\ll1$, at the two-loop order, showing that the gauge symmetry of the model is spontaneously broken, generating masses for the matter and gauge superfields via the Coleman--Weinberg mechanism, while supersymmetry remains manifest. This result is in agreement with the commutative versions of the present model~\cite{Ferrari:2010ex,Lehum:2010tt}.

An interesting issue is the presence of a term containing a factor of $1/\Theta$, which has a singularity in the commutative limit, $\Theta\rightarrow0$, revealing a type of UV/IR mixing. The presence of a such term seems to be intrinsic to vacuum diagrams used to evaluate the effective superpotential and is also present in the three-dimensional Wess--Zumino model~\cite{Ferrari:2009zx,Lehum:2010sh}.

Noncommutative non-Abelian extensions of present work should share the same properties of the noncommutative Abelian model studied here. One interesting question is what about more supersymmetric (e.g., ${\cal{N}}=2$) versions of this model? In fact, such work is currently in progress. Another possible extension would be searching for supersymmetry breaking, using techniques developed by Helayel-Neto et al~\cite{HelayelNeto:1984iv}, both in commutative and noncommutative versions of the supersymmetric Chern--Simons-matter model. 

\vspace{1cm}

\textbf{Acknowledgments}
\vspace{.5cm}

This work was partially supported by the Brazilian agency Conselho Nacional de Desenvolvimento Cient\'{\i}fico e Tecnol\'{o}gico (CNPq) by the project No. 303392/2010-0.

\appendix

\section{Noncommutative vertices}\label{vertex}

The noncommutative vertices are characterized by the presence of noncommutative phases. In this appendix, we write the important vertices to evaluate the diagrams drawn in the Figs. \ref{Fig1} and \ref{Fig2}. We do not consider the Faddeev--Popov vertices and diagrams involving Faddeev--Popov ghosts because they decouple from the other fields in our choice of gauge, $\xi=0$. The index of the vertex is related to the label of the vertex picture drawn in Fig. \ref{Fig3}. In momentum space, the noncommutative vertices can be written as    
\begin{eqnarray}\label{a01}
V_a=-\frac{\lambda}{4}~e^{-i[k_2\wedge(k_3+k_4)+k_3\wedge k_4]}\Phi_1(k_1)\Phi_1(k_2)\Phi_1(k_3)\Phi_1(k_4)~,
\end{eqnarray}
\begin{eqnarray}\label{a02}
V_b=-\frac{\lambda}{4}~e^{-i[k_2\wedge(k_3+k_4)+k_3\wedge k_4]}\Phi_2(k_1)\Phi_2(k_2)\Phi_2(k_3)\Phi_2(k_4)~,
\end{eqnarray}
\begin{eqnarray}\label{a03}
V_c=\frac{\lambda}{2}~e^{ik_4\wedge(k_2+k_3)}\left[2i\sin{(k_2\wedge k_3)}-e^{-ik_2\wedge k_3}\right]\Phi_1(k_1)\Phi_1(k_2)\Phi_2(k_3)\Phi_2(k_4)~,
\end{eqnarray}
\begin{eqnarray}\label{a04}
V_d=-\frac{g^2}{4}~e^{-i[k_2\wedge(k_3+k_4)+k_3\wedge k_4]} \Phi_1(k_1)\Phi_1(k_2)\Gamma^{\alpha}(k_3)\Gamma_{\alpha}(k_4)~,
\end{eqnarray}
\begin{eqnarray}\label{a05}
V_e=-\frac{g^2}{4}~e^{-i[k_2\wedge(k_3+k_4)+k_3\wedge k_4]} \Phi_2(k_1)\Phi_2(k_2)\Gamma^{\alpha}(k_3)\Gamma_{\alpha}(k_4)~,
\end{eqnarray}
\begin{eqnarray}\label{a06}
V_f=-\frac{g}{2}~\sin{(k_3\wedge k_2)} \Phi_1(k_1)D^{\alpha}\Phi_1(k_2)\Gamma_{\alpha}(k_3)~,
\end{eqnarray}
\begin{eqnarray}\label{a07}
V_g=-\frac{g}{2}~\sin{(k_3\wedge k_2)} \Phi_2(k_1)D^{\alpha}\Phi_2(k_2)\Gamma_{\alpha}(k_3)~,
\end{eqnarray}
\begin{eqnarray}\label{a08}
V_h=-\frac{g}{2}~\cos{(k_2\wedge k_3)} \left[\Phi_2(k_2)D^{\alpha}\Phi_1(k_1)\Gamma_{\alpha}(k_3)-\Phi_1(k_1)D^{\alpha}\Phi_2(k_2)\Gamma_{\alpha}(k_3)\right]~,
\end{eqnarray}
\begin{eqnarray}\label{a09}
V_i=-\lambda\varphi~\mathrm{e}^{-i~k_2\wedge k_3} \Phi_1(k_1)\Phi_1(k_2)\Phi_1(k_3)~,
\end{eqnarray}
\begin{eqnarray}\label{a10}
V_j=-\lambda\varphi~\mathrm{e}^{-i~k_2\wedge k_3} \Phi_2(k_1)\Phi_2(k_2)\Phi_1(k_3)~,
\end{eqnarray}
\begin{eqnarray}\label{a11}
V_k=-\frac{g^2}{2}\varphi~\mathrm{e}^{-i~k_2\wedge k_3} \Phi_1(k_1)\Gamma^{\alpha}(k_2)\Gamma_{\alpha}(k_3)~,
\end{eqnarray}
\begin{eqnarray}\label{a12}
V_l=-\frac{g}{3}\sin(k_3\wedge k_2)~\Gamma^{\alpha}(k_1)\Gamma^{\beta}(k_2)D_{\beta}\Gamma_{\alpha}(k_3)~,
\end{eqnarray}
\begin{eqnarray}\label{a13}
V_m=\frac{g^2}{6}~\sin{(k_4\wedge k_3)}\sin{[k_2\wedge(k_3+k_4)]}~
\Gamma^{\alpha}(k_1)\Gamma^{\beta}(k_2)\Gamma_{\beta}(k_3)\Gamma_{\alpha}(k_4)~.
\end{eqnarray}

\section{Evaluation of the Feynman graphs}\label{twoloop}

\begin{figure}[ht]
\includegraphics[width={13cm}]{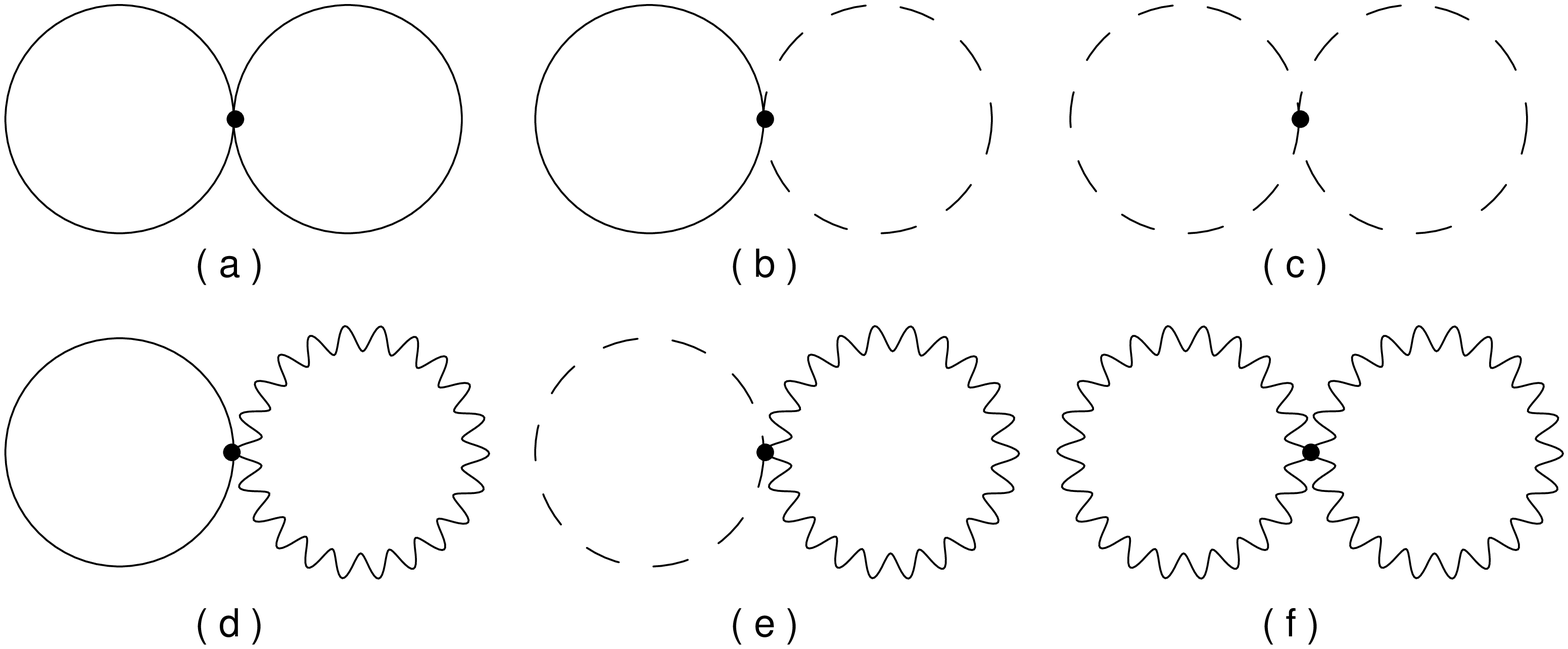}
\caption{UV finite two-loop diagrams which contribute to the effective action. Continuous lines represent the $\Phi_1$ propagator, dashed lines the $\Phi_2$ propagator and wavy lines the gauge superpotential propagator.}
\label{Fig1}
\end{figure}

The UV-finite Feynman diagrams which contribute to the two-loop order of the effective action are drawn in Fig. \ref{Fig1}. To evaluate the D-algebra of two-loop diagrams, we have used SusyMath~\cite{Ferrari:2007sc}. These contributions are given by
\begin{eqnarray}\label{b01}
\Gamma_{\ref{Fig1}a}=\frac{\lambda}{4}\int{d^5z}\int{\frac{d^3k}{(2\pi)^3}\frac{d^3q}{(2\pi)^3}} 
\frac{2+e^{-2ik\wedge q}}{(k^2+M_1^2)(q^2+M_1^2)}~,
\end{eqnarray}
\begin{eqnarray}\label{b02}
\Gamma_{\ref{Fig1}b}=\frac{\lambda}{2}\int{d^5z}\int{\frac{d^3k}{(2\pi)^3}\frac{d^3q}{(2\pi)^3}} 
\frac{e^{2iq\wedge k}}{(k^2+M_1^2)(q^2+M_1^2)}~,
\end{eqnarray}
\begin{eqnarray}\label{b03}
\Gamma_{\ref{Fig1}c}=\frac{\lambda}{4}\int{d^5z}\int{\frac{d^3k}{(2\pi)^3}\frac{d^3q}{(2\pi)^3}} 
\frac{2+e^{-2ik\wedge q}}{(k^2+M_2^2)(q^2+M_2^2)}~,
\end{eqnarray}
\begin{eqnarray}\label{b04}
\Gamma_{\ref{Fig1}d}=\frac{g^2}{4}\int{d^5z}\int{\frac{d^3k}{(2\pi)^3}\frac{d^3q}{(2\pi)^3}} 
\frac{1}{(k^2+M_1^2)(q^2+M_A^2)}~,
\end{eqnarray}
\begin{eqnarray}\label{b05}
\Gamma_{\ref{Fig1}e}=\frac{g^2}{4}\int{d^5z}\int{\frac{d^3k}{(2\pi)^3}\frac{d^3q}{(2\pi)^3}} 
\frac{1}{(k^2+M_2^2)(q^2+M_A^2)}~,
\end{eqnarray}
\begin{eqnarray}\label{b06}
\Gamma_{\ref{Fig1}f}=\frac{g^2}{2}\int{d^5z}\int{\frac{d^3k}{(2\pi)^3}\frac{d^3q}{(2\pi)^3}} 
\frac{\sin^2{(q\wedge k)}}{(k^2+M_A^2)(q^2+M_A^2)}~.
\end{eqnarray}

\begin{figure}[ht]
\includegraphics[width={10cm}]{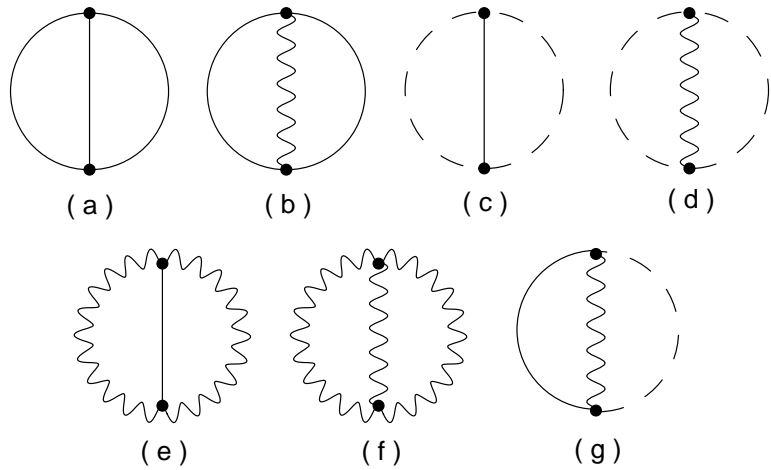}
\caption{UV logarithmically divergent graphs. These diagrams are the ones responsible for introducing the mass scale which spontaneously breaks the gauge invariance of the model.}   \label{Fig2}
\end{figure}

\begin{figure}[ht]
\includegraphics[width={13cm}]{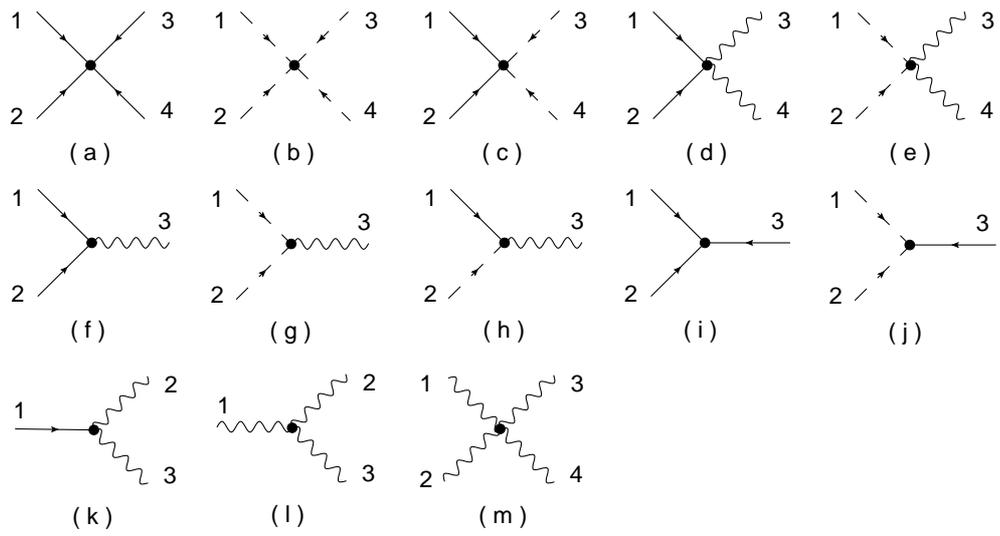}
\caption{Noncommutative vertices.}   
\label{Fig3}
\end{figure}

The contribution to the effective action which comes from the logarithmically divergent diagrams is
\begin{eqnarray}\label{c01}
\Gamma_{\ref{Fig2}a}=27\lambda^2 \int{d^5z}~\varphi^4\int{\frac{d^3k}{(2\pi)^3}\frac{d^3q}{(2\pi)^3}} 
\frac{e^{-ik\wedge q}\cos(k\wedge q)}{(k^2+M_1^2)(q^2+M_1^2)[(k+q)^2+M_1^2]}~,
\end{eqnarray}
\begin{eqnarray}\label{c02}
\Gamma_{\ref{Fig2}b}=-\frac{3g^2}{16} \int{d^5z}~\varphi^4\int{\frac{d^3k}{(2\pi)^3}\frac{d^3q}{(2\pi)^3}} 
\frac{\sin^2(q\wedge k)(12\lambda^2+\lambda g^2)}{(k^2+M_1^2)(q^2+M_A^2)[(k+q)^2+M_1^2]}~,
\end{eqnarray}
\begin{eqnarray}\label{c03}
\Gamma_{\ref{Fig2}c}=3\lambda^2 \int{d^5z}~\varphi^4\int{\frac{d^3k}{(2\pi)^3}\frac{d^3q}{(2\pi)^3}} 
\frac{e^{-ik\wedge q}\cos(k\wedge q)}{(k^2+M_2^2)(q^2+M_2^2)[(k+q)^2+M_1^2]}~,
\end{eqnarray}
\begin{eqnarray}\label{c04}
\Gamma_{\ref{Fig2}d}=-\frac{g^2}{8} \int{d^5z}~\varphi^4\int{\frac{d^3k}{(2\pi)^3}\frac{d^3q}{(2\pi)^3}} 
\frac{\sin^2(q\wedge k)(2\lambda^2+\lambda g^2)}{(k^2+M_2^2)(q^2+M_A^2)[(k+q)^2+M_2^2]}~,
\end{eqnarray}
\begin{eqnarray}\label{c05}
\Gamma_{\ref{Fig2}e}=&&-\frac{g^2}{8} \int{d^5z}~\varphi^2\int{\frac{d^3k}{(2\pi)^3}\frac{d^3q}{(2\pi)^3}} 
\frac{\mathrm{e}^{-i~k\wedge q}\cos(k\wedge q)}{(k^2+M_A^2)(q^2+M_A^2)[(k+q)^2+M_1^2]}\nonumber\\
&& \Big[2M_A+M_1+M_1M_A^2\frac{k\cdot q}{k^2~q^2} + M_A(k\cdot q)\left(\frac{1}{k^2}+\frac{1}{q^2}\right)\Big]~,
\end{eqnarray}
\begin{eqnarray}\label{c06}
\Gamma_{\ref{Fig2}f}=&&-\frac{g^2}{72} \int{d^5z}~\int{\frac{d^3k}{(2\pi)^3}\frac{d^3q}{(2\pi)^3}} 
\dfrac{\sin^2(k\wedge q)}{(k^2+M_A^2)(q^2+M_A^2)[(k+q)^2+M_A^2]}\nonumber\\
&&\Big{\{} \frac{4M_A^2q^2}{(k+q)^2}+\frac{3M_A^2q^2(k\cdot q)}{k^2(k+q)^2}+\frac{7M_A^2(k\cdot q)}{k^2}
-\frac{M_A^2}{(k+q)^2}[(k\cdot q)+k^2]\nonumber\\
&&-2(k\cdot q)-2(q^2+M_A^2)-M_A^2\Big{\}}~,
\end{eqnarray}
\begin{eqnarray}\label{c07}
\Gamma_{\ref{Fig2}g}=&&-\frac{g^2}{8} \int{d^5z}~\int{\frac{d^3k}{(2\pi)^3}\frac{d^3q}{(2\pi)^3}} 
\dfrac{\cos^2(k\wedge q)}{(k^2+M_1^2)(q^2+M_2^2)[(k+q)^2+M_A^2]}\nonumber\\
&&\Big{\{}\frac{M_A}{(k+q)^2}[(3M_1-2M_2)q^2+(M_1-3M_2)k^2]+4M_A\frac{k\cdot q}{(k+q)^2}(M_2-M_1)\nonumber\\
&&+4M_1M_2+M_A(M_1+M_2)+2(k^2+q^2)\Big{\}}~.
\end{eqnarray}

Considering the noncommutativity matrix $\Theta_{\mu\nu}=\epsilon_{0\mu\nu}\Theta$, in the limit of small noncommutativity, all diagrams result in similar integrals to those evaluated in Ref.~\cite{Ferrari:2009zx}. Summing up all contributions, i.e., tree, one-loop, and two-loop contributions, the K\"alerian effective superpotential can be written as  
\begin{eqnarray}\label{d01}
K_{eff}=\left[\frac{\lambda}{4}+f(\lambda,g,\epsilon)\right]\varphi^4+e\varphi^4\ln{\frac{\varphi^2}{\mu}}+\frac{h(g,\lambda)}{\Theta^2\varphi^4}+C\varphi^4+\mathcal{O}(\Theta)~,
\end{eqnarray}

\noindent
where $\mu$ is a mass scale introduced by the renormalization by dimensional reduction~\cite{Siegel:1979wq}, $C$ is a counterterm, $f(\lambda,g,1/\epsilon)$ is a function of the coupling constants, and $\epsilon=(D-3)$, $h(g,\lambda)$ is a function of the coupling constants, $e=a_1~g^6+a_2~g^4\lambda+a_2~g^4\lambda+a_3~g^2\lambda^2+a_4~\lambda^3$, with the $a_i$ numerical factors.

\end{document}